\documentclass[12pt]{article}
\usepackage[english]{babel}
\usepackage{amsmath,amsthm,amssymb,amsfonts,url}
\usepackage{mathptmx}      % use Times fonts if available on your TeX system

\usepackage{color}
\newtheorem{theorem}{Theorem}

\newtheorem{remark}{Remark}
\newtheorem{example}{Example}

\begin{document}

\title{Nonlinear first order partial differential equations reducible to first order homogeneous and autonomous quasilinear ones}

\author{M.~Gorgone and F.~Oliveri\\
\ \\
{\footnotesize MIFT Department, University of Messina}\\
{\footnotesize Viale F. Stagno d'Alcontres 31, 98166 Messina, Italy}\\
{\footnotesize mgorgone@unime.it; foliveri@unime.it}
}

\date{Published in \textit{Ricerche Mat.} \textbf{66}, 51--63 (2017).}

\maketitle

\begin{abstract}
A theorem providing necessary conditions enabling one to map a nonlinear system of first order partial differential equations to an equivalent first order autonomous and homogeneous quasilinear system is given. The reduction to quasilinear form is performed by constructing the canonical variables associated to the Lie point symmetries admitted by the nonlinear system. Some applications to relevant partial differential equations are given. 
\end{abstract}


%\subclass{35A30 \and 58J70 \and 58J72}

\section{Introduction}
\label{sec:introduction}
Lie group analysis 
\cite{Ovsiannikov,Ibragimov,Olver,Baumann,BlumanAnco,BlumanCheviakovAnco2009}
provide a unified and elegant algorithmic framework for a deep understanding and fruitful handling 
of differential equations. It is known that Lie point symmetries admitted by ordinary differential equations
allow for their order lowering; on the contrary, in the case of partial differential equations the symmetries can be used for the determination of special (invariant) solutions of initial and boundary value problems. Also, the Lie symmetries are important ingredients in the derivation of conserved quantities, or in the construction of relations between different differential equations that turn out to be equivalent 
\cite{BlumanAnco,BlumanCheviakovAnco2009,DonatoOliveri1994,DonatoOliveri1995,%
DonatoOliveri1996,CurroOliveri2008,Oliveri2010,Oliveri2012}. The use of Lie point symmetries is especially useful when we have to deal with nonlinear partial differential equations. 

In this paper, we shall consider nonlinear systems of first order partial differential equations; 
nevertheless, it is worth of being recalled that higher order partial differential equations can always 
be rewritten (though not in a unique way!) as first order systems.

Among the first order systems of partial differential equations, a special role is played by quasilinear systems either for their mathematical properties or for their ubiquity in the applications.
In fact, many physical problems may be mathematically modeled  
in terms of first order balance laws, say 
\begin{equation}
\label{balance}
\sum_{i=1}^n\frac{\partial\mathbf{F}^i(\mathbf{u})}{\partial x_i}=\mathbf{G}(\mathbf{u}),
\end{equation}
where $\mathbf{u}\in \mathbb{R}^m$ denotes the unknown vector field, $\mathbf{x}\in \mathbb{R}^n$
the set of independent variables, 
$\mathbf{F}^i(\mathbf{u})$ the components of a flux, and $\mathbf{G}(\mathbf{u})$ the production 
term (for dynamical systems the first component $x_1$ of the independent variables is the time, and the components of $\mathbf{F}^1$ are the densities of some physical quantities); when $\mathbf{G}(\mathbf{u})\equiv\mathbf{0}$, we have a system of conservation laws.
Systems like (\ref{balance}) fall in the more general class of nonhomogeneous
quasilinear first order systems of partial differential equations,
\begin{equation}
\label{quasilinear_auto}
\sum_{i=1}^n A^i(\mathbf{u})\frac{\partial\mathbf{u}}{\partial x_i}=\mathbf{G}(\mathbf{u}),
\end{equation}
where $A^i$ ($i=1,\ldots,n$) are $m\times m$ matrices with entries 
depending on the field $\mathbf{u}$.

Special problems may require to consider systems where the coefficients may also depend on the 
independent variables $\mathbf{x}$, accounting for material inhomogeneities, or 
special geometric assumptions, or external actions, so that one has to consider, in general, 
nonautonomous and/or nonhomogeneous quasilinear systems of the form
\begin{equation}
\label{quasilinear_nonauto}
\sum_{i=1}^n A^i(\mathbf{x},\mathbf{u})\frac{\partial\mathbf{u}}{\partial x_i}=
\mathbf{G}(\mathbf{x},\mathbf{u}).
\end{equation}

In \cite{Oliveri2012} it has been proved a theorem establishing necessary and sufficient conditions in order to map a system like (\ref{quasilinear_nonauto}) into an autonomous and homogeneous quasilinear system under the action of a one-to-one point variable transformation like
\begin{equation}
\mathbf{z}=\mathbf{Z}(\mathbf{x}), \qquad \mathbf{w}=\mathbf{W}(\mathbf{x},\mathbf{u}).
\end{equation}

The possibility of reducing system~(\ref{quasilinear_nonauto}) to autonomous 
and homogeneous form is intimately related to the
symmetry properties of the model under investigation \cite{Oliveri2012}; 
remarkably, when this approach is applicable, it is possible to construct explicitly the 
map transforming nonhomogeneous and/or nonautonomous  quasilinear systems to
homogeneous and autonomous form. The key idea is
that any  homogeneous and autonomous first order quasilinear
system is invariant with respect to $n$ independent translations of the independent variables and with respect to a uniform scaling of all independent variables. These symmetries span an $(n+1)$--dimensional solvable Lie algebra containing an $n$--dimensional Abelian Lie subalgebra. 

In this paper, we consider systems of first order \emph{nonlinear} partial differential equations with the aim of investigating whether they can be reduced by an invertible point transformation to an equivalent first order system of autonomous and homogeneous quasilinear equations. First order homogeneous and autonomous quasilinear systems
possess many relevant features: for instance, they admit self--similar solutions suitable to describe rarefaction waves. Also, many efficient numerical schemes useful for investigating physically relevant problems are available for such a kind of systems. 
Here we prove a theorem giving necessary conditions  for the transformation of a nonlinear first order system of partial differential equations to  autonomous and homogeneous quasilinear form.
Some examples concerned with the first order systems related to second order Monge--Amp\`ere equations in
$(1+1)$, $(2+1)$ and $(3+1)$ dimensions where the procedure works are discussed.

\section{General nonlinear systems}
Let us consider a general first order nonlinear system of PDEs,
\begin{equation}
\label{general}
\Delta\left(\mathbf{x},\mathbf{u},\mathbf{u}^{(1)}\right)=0,
\end{equation}
where $\mathbf{x}\equiv(x_1,\ldots,x_n)$ are the independent variables, 
$\mathbf{u}\equiv(u_1(\mathbf{x}),\ldots,u_m(\mathbf{x}))$ the dependent variables, and $\mathbf{u}^{(1)}$ denotes the set of all first order derivatives of the $u_i$'s with respect to the $x_j$'s.
We want to exploit the possibility of constructing an invertible mapping like
\begin{equation}
\label{transf}
\mathbf{z}=\mathbf{Z}(\mathbf{x},\mathbf{u}), \qquad \mathbf{w}=\mathbf{W}(\mathbf{x},\mathbf{u}) 
\end{equation} 
($\mathbf{z}\equiv(z_1,\ldots,z_n)$, $\mathbf{w}\equiv(w_1,\ldots,w_m)$), allowing us to map it  to a quasilinear homogeneous and autonomous system. 
When this is possible then \emph{necessarily} the nonlinear system has to possess a suitable 
$(n+1)$--dimensional solvable Lie algebra as subalgebra of the algebra of its Lie point symmetries.  

Let us suppose that system (\ref{general}) can be mapped through an invertible point transformation like (\ref{transf}) into an autonomous and homogeneous quasilinear system, say
\begin{equation}
\label{quasilinear}
\sum_{i=1}^n A^i(\mathbf{w})\frac{\partial\mathbf{w}}{\partial z_i}=\mathbf{0},
\end{equation}
where $A^i$ ($i=1,\ldots,n$) are $m\times m$ matrices with entries 
depending at most on $\mathbf{w}$. Every system like (\ref{quasilinear}) admits the Lie symmetries
generated by the following vector fields:
\begin{equation}
\Xi_i=\frac{\partial}{\partial z_i}, \quad (i=1,\ldots,n), \qquad \Xi_{n+1}=\sum_{i=1}^nz_i\frac{\partial}{\partial z_i};
\end{equation}
these vector fields  span an $(n+1)$--dimensional solvable Lie
algebra where the only non--zero commutators are
\begin{equation}
[\Xi_i,\Xi_{n+1}]=\Xi_i, \qquad i=1,\ldots,n.
\end{equation}
Moreover, the first $n$ vector fields span an $n$--dimensional Abelian Lie algebra, and generate a distribution of rank $n$.

Since neither the rank of a distribution nor the Lie bracket of two infinitesimal generators of
symmetries is affected by an invertible change of coordinates, it follows that if a system of the
form (\ref{general}) can be mapped by (\ref{transf}) to the form (\ref{quasilinear}), it has to admit,
as subalgebra of the Lie algebra of its point symmetries, an $(n+1)$--dimensional
Lie algebra with a suitable algebraic structure.

Therefore, the following theorem is proved.
\begin{theorem}\label{th_1}
A necessary condition in order the nonlinear system
\begin{equation}
\Delta\left(\mathbf{x},\mathbf{u},\mathbf{u}^{(1)}\right)=0
\end{equation}
be transformed by the invertible map
\begin{equation}
\mathbf{z}=\mathbf{Z}(\mathbf{x},\mathbf{u}), \qquad \mathbf{w}=\mathbf{W}(\mathbf{x},\mathbf{u}) 
\end{equation}
into an autonomous and homogeneous first order quasilinear system is that
it admits as subalgebra of its Lie point symmetries an $(n+1)$--dimensional Lie algebra 
spanned by
\begin{equation}
\Xi_i=\sum_{j=1}^n\xi_i^j(\mathbf{x},\mathbf{u})\frac{\partial}{\partial x_j}+\sum_{A=1}^m\eta_i^A(\mathbf{x},\mathbf{u})\frac{\partial}{\partial u_A},\;\;(i=1,\ldots, n+1),
\end{equation}
such that
\begin{equation}
\left[\Xi_i,\Xi_j\right]=0, \qquad \left[\Xi_i,\Xi_{n+1}\right]=\Xi_i, \qquad i,j=1,\ldots,n.
\end{equation}
Furthermore, the vector fields $\Xi_1,\ldots,\Xi_n$ have to generate a distribution of rank $n$.
The new independent and dependent variables are the canonical variables associated to the 
symmetries generated by $\Xi_1,\ldots,\Xi_n$, say
\begin{equation}
\Xi_i(z_j)=\delta_{ij}, \qquad \Xi_i(\mathbf{w})=\mathbf{0}, \qquad i,j=1,\ldots,n,
\end{equation}
where $\delta_{ij}$ is the Kronecker symbol. Finally, the
variables $\mathbf{w}$, which by construction are invariants of $\Xi_1,\ldots,\Xi_n$, 
must result invariant with respect to $\Xi_{n+1}$ too.
\end{theorem}

\begin{remark}
For the details of the construction of the new independent and dependent variables one can refer to \cite{Oliveri2010,Oliveri2012}. 
\end{remark}

\begin{remark}
The conditions required by Theorem~\ref{th_1} are not sufficient to guarantee the transformation to
quasilinear form. Nevertheless, if the nonlinear system of partial differential equations involves the derivatives in polynomial form, then those conditions are necessary and sufficient for the mapping  
into a system where each equation is a homogeneous polynomial in the derivatives. 
\end{remark}

\begin{remark}
Theorem~\ref{th_1} can be used also when the nonlinear source system is autonomous. In such a way, when the hypotheses of the theorem are satisfied, the target system should be autonomous too; in fact, only in this case the invariance with respect to the homogeneous scaling of the independent variables of the target system implies that the system
is a homogeneous polynomial in the derivatives (a quasilinear system if the degree of the homogeneous polynomial is 1),    
\end{remark}

\section{Applications}
In this Section, we provide some examples of  systems of first order nonlinear partial differential equations, whose Lie symmetries satisfy the conditions of Theorem~\ref{th_1}, and prove that they can be transformed 
under suitable conditions to quasilinear autonomous and homogeneous systems. The nonlinear first order 
systems are obtained from second order $(1+1)$--, $(2+1)$-- and $(3+1)$--dimensional Monge--Amp\`ere 
equations. It is worth of noticing that some classes of these systems have been proved to be linearizable by 
invertible point transformations in \cite{Oliveri1998}.

Hereafter, to shorten the formulas, we use the notation $u_{,i}$ and $u_{,ij}$ to indicate the first order partial derivatives of $u$ with respect to $x_i$, and the second order partial derivatives of $u$ with respect to $x_i$ and $x_j$, respectively. Moreover, we shall denote with $f_{;i}$ and $f_{;ij}$ the first order partial derivative of the function $f$ with respect to $u_i$ and the second order partial derivatives of $f$ with respect to $u_i$ and $u_j$, respectively. 
 
\subsection{Monge--Amp\`ere equation in $(1+1)$ dimensions}
In 1968, Boillat \cite{Boillat} proved that the most general completely exceptional second order 
equation in $(1+1)$ dimensions is the well known  Monge--Amp\`ere equation,
\begin{equation}
\label{monge1+1}
\kappa_1\left(u_{,11}u_{,22}-u_{,12}^2\right)+\kappa_2u_{,11}+\kappa_3u_{,12}+\kappa_4u_{,22}+\kappa_5=0,
\end{equation}
with $u(x_1,x_2)$ a scalar function, and $\kappa_i\left(x_1,x_2,u,u_{,1}, u_{,2}\right)$ $(i=1,\ldots ,5)$ arbitrary smooth functions of the indicated arguments.

By means of the positions
\begin{equation}
u_1=u_{,1}, \qquad u_2=u_{,2},
\end{equation}
along with the assumptions that the functions $\kappa_i$ $(i=1,\ldots,5)$ depend at most on $(u_1,u_2)$, we obtain the following nonlinear first order system:
\begin{equation}\label{monge1+1_equiv}
\begin{aligned}
&u_{2,1}-u_{1,2}=0, \\
&\kappa_1\left(u_{1,1}u_{2,2}-u_{1,2}^2\right)+\kappa_2u_{1,1}+\kappa_3u_{1,2}+\kappa_4u_{2,2}+\kappa_5=0.
\end{aligned}
\end{equation}
Through the substitutions
\begin{equation}
u_1\rightarrow u_1+\alpha_1 x_1+\alpha_2 x_2, \qquad u_2\rightarrow u_2+\alpha_2 x_1+\alpha_3 x_2,  
\end{equation}
$\alpha_i$ $(i=1,\ldots,3)$ being constant, we get the system
\begin{equation}\label{monge1+1_equivbis}
\begin{aligned}
&u_{2,1}-u_{1,2}=0, \\
&\kappa_1\left(u_{1,1}u_{2,2}-u_{1,2}^2\right)+(\alpha_3\kappa_1+\kappa_2)u_{1,1}+(-2\alpha_2\kappa_1+\kappa_3)u_{1,2}+(\alpha_1\kappa_1+\kappa_4)u_{2,2}\\
&\qquad+((\alpha_1\alpha_3-\alpha_2^2)\kappa_1+\alpha_1\kappa_2+\alpha_2\kappa_3+\alpha_3\kappa_4+\kappa_5)=0.
\end{aligned}
\end{equation}
The nonlinear system (\ref{monge1+1_equivbis}) becomes homogeneous if
\begin{equation}
\kappa_5=-((\alpha_1\alpha_3-\alpha_2^2)\kappa_1+\alpha_1\kappa_2+\alpha_2\kappa_3+\alpha_3\kappa_4),
\end{equation}
and in such a case it is straightforward to recognize that it 
admits the Lie point symmetries spanned by the operators
\begin{equation}
\Xi_1=\frac{\partial}{\partial x_1},\qquad
\Xi_2=\frac{\partial}{\partial x_2},\qquad
\Xi_3=\left(x_1-f_{;1}\right)\frac{\partial}{\partial x_1}+\left(x_2-f_{;2}\right)\frac{\partial}{\partial x_2},
\end{equation}
where $f(u_1,u_2)$ is an arbitrary smooth function of its arguments, provided that 
\begin{equation}
\kappa_1=\frac{-\kappa_2f_{;22}+\kappa_3f_{;12}-\kappa_4f_{;11}}{1+\alpha_3 f_{;22}+2\alpha_2 f_{;12}+\alpha_1 f_{;11}}.
\end{equation}

Since
\begin{equation}
\left[\Xi_1,\Xi_2\right]=0,\qquad \left[\Xi_1,\Xi_3\right]=\Xi_1,\qquad \left[\Xi_2,\Xi_3\right]=\Xi_2,
\end{equation}
we introduce the new variables
\begin{equation}
\label{newvars11}
z_1=x_1-f_{;1},\qquad z_2=x_2-f_{;2},\qquad w_1=u_1, \qquad w_2=u_2,
\end{equation}
and the generators of the point symmetries write as
\begin{equation}
\Xi_1=\frac{\partial}{\partial z_1}, \qquad \Xi_2=\frac{\partial}{\partial z_2}, \qquad
\Xi_3=z_1\frac{\partial}{\partial z_1}+z_2\frac{\partial}{\partial z_2}.
\end{equation}
In terms of the new variables (\ref{newvars11}),  the nonlinear system (\ref{monge1+1_equivbis}) 
becomes
\begin{equation}
\begin{aligned}
& w_{2,1}-w_{1,2}=0, \\
&(\alpha_3\kappa_1+\kappa_2)w_{1,1}+(-2\alpha_2\kappa_1+\kappa_3)w_{1,2}+(\alpha_1\kappa_1+\kappa_4)w_{2,2}=0,
\end{aligned}
\end{equation}
\emph{i.e.}, reads as an autonomous and homogeneous quasilinear system.

The following example provides a physical system leading to a Monge--Amp\`ere equation.

\begin{example}[One--dimensional Euler equations of isentropic fluids]
Let us consider the Euler equations for an isentropic fluid
\begin{equation}
\begin{aligned}
&\rho_{,1}+(\rho v)_{,2}=0,\\
&(\rho v)_{,1}+(\rho v^2+p(\rho,s))_{,2}=0,\\
& s_{,1}+v s_{,2}=0,
\end{aligned}
\end{equation}
where $\rho$ is the fluid density, $v$ the velocity, $s$ the entropy, $p(\rho,s)$ the pressure which is a function of the density and the entropy, $x_1$ the time, and $x_2$ the space variable.

By introducing $\phi$ such that
\begin{equation}
\rho=\phi_{,2},\qquad \rho v=-\phi_{,1}
\end{equation}
it results $s=s(\phi)$. Moreover, by introducing $\psi$ and $u$ such that
\begin{equation}
\begin{aligned}
&\rho v=\psi_{,2}=-\phi_{,1}, \qquad &&\rho v^2+p=-\psi_{,1},\\
&\psi=-u_{,1}, &&\phi=u_{,2},
\end{aligned}
\end{equation}
we arrive to the nonlinear equation
\begin{equation}
u_{,11}=\frac{u_{,12}^2}{u_{,22}}+p(u_{,22},s(u_{,1})).
\end{equation}
This equation becomes of Monge--Amp\`ere type for the class of fluids characterized by the constitutive law 
of Von K\'arm\'an \cite{VonKarman}
\begin{equation}
p=-\frac{\kappa^2(s)}{\rho}+b(s),
\end{equation}
where $\kappa^2(s)>0$ and $b(s)$ are functions of the entropy. What we get is
\begin{equation}
\label{mongevonkarman}
u_{,11}u_{,22}-u_{,12}^2+\kappa^2(s(u_{,2}))-b(s(u_{,2}))u_{,22}=0.
\end{equation}
The nonlinear first order system derived from equation~(\ref{mongevonkarman}) belongs to the class of
equations (\ref{monge1+1_equiv}) and is mapped to a homogeneous and autonomous quasilinear system provided that
\begin{equation}
\begin{aligned}
&\kappa^2(s(u_2))=\alpha_2^2-\alpha_3(\alpha_1+b(s(u_2)),\\
&b(s(u_2))=\frac{1+\alpha_3 f_{;22}+2\alpha_2 f_{;12}+\alpha_1 f_{;11}}{f_{;11}},
\end{aligned}
\end{equation}
and $f(u_1,u_2)$ is such that
\begin{equation}
\frac{\partial}{\partial u_1}\left(\frac{1+\alpha_3 f_{;22}+2\alpha_2 f_{;12}}{f_{;11}}\right)=0.
\end{equation}
\end{example}

\subsection{Monge--Amp\`ere equation in $(2+1)$ dimensions}
The most general second order hyperbolic completely exceptional equation in $(2+1)$ dimensions has been determined in 1973 by Ruggeri \cite{Ruggeri}; it is a linear combination of the determinant and all minors extracted from the $3\times 3$ Hessian 
matrix of $u(x_1,x_2,x_3)$ with coefficients $\kappa_i$ $(i=1,\ldots,14)$ depending on the independent variables, the dependent variable and its first order derivatives. This equation can be written in the following form:
\begin{equation}\label{monge2+1}
\begin{aligned}
&\kappa_1 H+\kappa_2 \frac{\partial H}{\partial u_{,11}}+\kappa_3 \frac{\partial H}{\partial u_{,12}}+
\kappa_4 \frac{\partial H}{\partial u_{,13}}+\kappa_5 \frac{\partial H}{\partial u_{,22}}+
\kappa_6 \frac{\partial H}{\partial u_{,23}}+\kappa_7\frac{\partial H}{\partial u_{,33}}\\
&+\kappa_8 u_{,11}+\kappa_9 u_{,12}+\kappa_{10} u_{,13}+\kappa_{11} u_{,22}+\kappa_{12} u_{,23}+\kappa_{13} u_{,33}+\kappa_{14}=0,
\end{aligned}
\end{equation}
where $H$ is the determinant of the $3\times 3$ Hessian matrix of $u$. 

Let us assume $\kappa_i$ $(i=1,\ldots,14)$ depending at most on first order derivatives of $u$.
By means of the positions
\begin{equation}
u_1=u_{,1}, \qquad u_2=u_{,2}, \qquad u_3=u_{,3},
\end{equation}
the following nonlinear first order system is obtained:
\begin{equation}\label{monge2+1_equiv}
\begin{aligned}
&u_{2,1}-u_{1,2}=0,\qquad u_{3,1}-u_{1,3}=0,\qquad u_{3,2}-u_{2,3}=0,\\
&\kappa_1 H+\kappa_2 \frac{\partial H}{\partial u_{1,1}}+\kappa_3 \frac{\partial H}{\partial u_{1,2}}+
\kappa_4 \frac{\partial H}{\partial u_{1,3}}+\kappa_5 \frac{\partial H}{\partial u_{2,2}}+
\kappa_6 \frac{\partial H}{\partial u_{2,3}}+\kappa_7 \frac{\partial H}{\partial u_{3,3}}\\
&+\kappa_8 u_{1,1}+\kappa_9 u_{1,2}+\kappa_{10} u_{1,3}+\kappa_{11} u_{2,2}+\kappa_{12} u_{2,3}+\kappa_{13} u_{3,3}+\kappa_{14}=0.
\end{aligned}
\end{equation}
As done in the previous subsection, the substitutions
\begin{equation}
\begin{aligned}
&u_1\rightarrow u_1+\alpha_1 x_1+\alpha_2 x_2+\alpha_3 x_3, \\
&u_2\rightarrow u_2+\alpha_2 x_1+\alpha_4 x_2+\alpha_5 x_3, \\
&u_3\rightarrow u_3+\alpha_3 x_1+\alpha_5 x_2+\alpha_6 x_3, 
\end{aligned}
\end{equation}
$\alpha_i$ $(i=1,\ldots,6)$ being constant, provided that
\begin{equation}
\begin{aligned}
\kappa_{14}&=(\alpha_1\alpha_5^2+\alpha_2^2\alpha_6-\alpha_1\alpha_4\alpha_6-2\alpha_2\alpha_3\alpha_5+\alpha_3^2\alpha_4)\kappa_1-(\alpha_5^2-\alpha_4\alpha_6)\kappa_2\\
&+2(\alpha_2\alpha_6-\alpha_3\alpha_5)\kappa_3+2(\alpha_3\alpha_4-\alpha_2\alpha_5)\kappa_4+(\alpha_3^2-\alpha_1\alpha_6)\kappa_5\\
&+2(\alpha_1\alpha_5-\alpha_2\alpha_3)\kappa_6+(\alpha_2^2-\alpha_1\alpha_4)\kappa_7-\alpha_1\kappa_8-\alpha_2\kappa_9-\alpha_3\kappa_{10}\\
&-\alpha_4\kappa_{11}-\alpha_5\kappa_{12}-\alpha_6\kappa_{13},
\end{aligned}
\end{equation} 
allows us to get a homogeneous system with the same differential structure as (\ref{monge2+1_equiv}), say
\begin{equation}\label{monge2+1_equivbis}
\begin{aligned}
&u_{2,1}-u_{1,2}=0,\qquad u_{3,1}-u_{1,3}=0,\qquad u_{3,2}-u_{2,3}=0,\\
&\widehat\kappa_1 H+\widehat\kappa_2 \frac{\partial H}{\partial u_{1,1}}+\widehat\kappa_3 \frac{\partial H}{\partial u_{1,2}}+
\widehat\kappa_4 \frac{\partial H}{\partial u_{1,3}}+\widehat\kappa_5 \frac{\partial H}{\partial u_{2,2}}+
\widehat\kappa_6 \frac{\partial H}{\partial u_{2,3}}+\widehat\kappa_7 \frac{\partial H}{\partial u_{3,3}}\\
&+\widehat\kappa_8 u_{1,1}+\widehat\kappa_9 u_{1,2}+\widehat\kappa_{10} u_{1,3}+\widehat\kappa_{11} u_{2,2}+\widehat\kappa_{12} u_{2,3}+\widehat\kappa_{13} u_{3,3}=0.
\end{aligned}
\end{equation}
where the expression of $\widehat\kappa_i$ in terms of the coefficients $\kappa_i$ $(i=1,\ldots,13)$ and the constants
$\alpha_j$ $(j=1,\ldots,6)$ can be easily found.

The nonlinear system (\ref{monge2+1_equivbis}) admits the Lie symmetries spanned by the operators
\begin{equation}
\begin{aligned}
&\Xi_1=\frac{\partial}{\partial x_1},\qquad
\Xi_2=\frac{\partial}{\partial x_2},\qquad \Xi_3=\frac{\partial}{\partial x_3},\\
&\Xi_4=\left(x_1-f_{;1}\right)\frac{\partial}{\partial x_1}+\left(x_2-f_{;2}\right)\frac{\partial}{\partial x_2}+\left(x_3-f_{;3}\right)\frac{\partial}{\partial x_3},
\end{aligned}
\end{equation}
where $f(u_1,u_2,u_3)$ is a smooth arbitrary function of its arguments,
provided that the following relations hold true:
\begin{equation}
\label{cond_2+1}
\begin{aligned}
&\widehat\kappa_{1} -H_f^{11}\widehat\kappa_{8}- H_f^{12}\widehat\kappa_{9}-H_f^{13}\widehat\kappa_{10} -H_f^{22}\widehat\kappa_{11}   -H_f^{23}\widehat\kappa_{12}-H_f^{33}\widehat\kappa_{13}=0,\\
&\widehat\kappa_{2}+ f_{;33}\widehat\kappa_{11} -f_{;23}\widehat\kappa_{12} + f_{;22}\widehat\kappa_{13} =0,\\
&2\widehat\kappa_{3} - f_{;33}\widehat\kappa_{9} + f_{;23}\widehat\kappa_{10}+ f_{;13}\widehat\kappa_{12} -2f_{;12}\kappa_{13}=0,\\
&2\widehat\kappa_{4}  + f_{;23}\widehat\kappa_{9} - f_{;22}\widehat\kappa_{10}- 2f_{;13}\widehat\kappa_{11}+ f_{;12}\widehat\kappa_{12} =0,\\
&\widehat\kappa_{5}+ f_{;33}\widehat\kappa_{8}-f_{;13}\widehat\kappa_{10} + f_{;11}\widehat\kappa_{13} =0,\\
&2\widehat\kappa_{6}- 2f_{;23}\widehat\kappa_{8}  + f_{;13}\widehat\kappa_{9} + f_{;12}\widehat\kappa_{10}- f_{;11}\widehat\kappa_{12} =0,\\
&\widehat\kappa_{7}+ f_{;22}\widehat\kappa_{8}-f_{;12}\widehat\kappa_{9} + f_{;11}\widehat\kappa_{11} =0,\\
\end{aligned}
\end{equation}
$H_f^{ij}$ denoting the cofactor of the $(i,j)$ entry of the Hessian matrix $H_f$ of the function
$f(u_1,u_2,u_3)$. It is evident that conditions (\ref{cond_2+1}) place severe restrictions on the coefficients of
system (\ref{monge2+1_equivbis}); in fact, they state that the functions $\widehat\kappa_i$ $(i=1,\ldots,7)$ have to be expressed
in terms of the coefficients $\widehat\kappa_i$ $(i=8,\ldots,13)$ and the function $f$. 
These symmetries generate a 4--dimensional solvable Lie algebra,
\begin{equation}
\left[\Xi_i,\Xi_j\right]=0,\qquad \left[\Xi_i,\Xi_4\right]=\Xi_i,
\qquad (i,j=1,2,3),
\end{equation}
whereupon we may introduce the new variables
\begin{equation}
\label{newvars21}
\begin{array}{lll}
z_1=x_1-f_{;1},\quad &z_2=x_2-f_{;2},\quad & z_3= x_3-f_{;3},\\ 
w_1=u_1,\quad  &w_2=u_2,\quad & w_3=u_3,
\end{array}
\end{equation}
and the generators of the point symmetries write as
\begin{equation}
\Xi_1=\frac{\partial}{\partial z_1}, \qquad \Xi_2=\frac{\partial}{\partial z_2}, \qquad
\Xi_3=\frac{\partial}{\partial z_3}, \qquad
\Xi_4=z_1\frac{\partial}{\partial z_1}+z_2\frac{\partial}{\partial z_2}
+z_3\frac{\partial}{\partial z_3}. 
\end{equation}

In terms of the new variables (\ref{newvars21}), the nonlinear system (\ref{monge2+1_equivbis}) reduces to
\begin{equation}
\begin{aligned}
&w_{2,1}-w_{1,2}=0,\qquad w_{3,1}-w_{1,3}=0,\qquad w_{3,2}-w_{2,3}=0,\\
&\widehat\kappa_8w_{1,1} + \widehat\kappa_9w_{1,2} + \widehat\kappa_{10}w_{1,3} + \widehat\kappa_{11}w_{2,2} + \widehat\kappa_{12}w_{2,3}+ \widehat\kappa_{13}w_{3,3}=0,
\end{aligned}
\end{equation}
\emph{i.e.}, an autonomous and homogeneous quasilinear system.

\subsection{Monge--Amp\`ere equation in $(3+1)$ dimensions}
The most general second order completely exceptional equation in $(3+1)$ dimensions 
has been characterized by Donato \emph{et al.} \cite{Donato} and once again it is given as a linear combination of the determinant and all minors extracted from the $4\times 4$ Hessian 
matrix of $u(x_1,x_2,x_3,x_4)$ with coefficients $\kappa_i$ $(i=1,\ldots,43)$ depending on the independent variables, the dependent variable and its first order derivatives:
\begin{equation}\label{monge3+1}
\begin{aligned}
&\kappa_1 H+\sum_{r}\kappa_r\frac{\partial H}{\partial u_{,ij}}
+\sum_{s}\kappa_s\frac{\partial^2 H}{\partial u_{,kl}\partial u_{,mn}}+\sum_{r}\kappa_{r+31}u_{,ij}+\kappa_{43}=0,\\
&i,j,k,l,m,n=1,\ldots,4,\quad i\leq j,\quad k<l,\quad k\leq m<n,\\
&r=\frac{i(9-i)}{2}+j-3,\quad s=\sigma_{mn}+\frac{\sigma_{kl}(13-\sigma_{kl})}{2}+6,\\
&\sigma_{ab}=4(a-1)-\frac{a(a+1)}{2}+b,
\end{aligned}
\end{equation}
where $H$ is the determinant of the $4\times 4$ Hessian matrix of $u$;
actually, the Monge--Amp\`ere equation in $(3+1)$ dimensions involves only $42$ independent coefficients because
\begin{equation}
\frac{\partial^2 H}{\partial u_{,12}\partial u_{,34}}+\frac{\partial^2 H}{\partial u_{,13}\partial u_{,24}}+
\frac{\partial^2 H}{\partial u_{,14}\partial u_{,23}}=0.
\end{equation}
Hereafter, we assume without loss of generality, $\kappa_{24}=0$, and the remaining functions $\kappa_i$ depending at most on first order derivatives.

By means of the positions
\begin{equation}
u_1=u_{,1}, \qquad u_2=u_{,2}, \qquad u_3=u_{,3},\qquad u_4=u_{,4},
\end{equation}
the following nonlinear first order system is obtained:
\begin{equation}\label{monge3+1_equiv}
\begin{aligned}
&u_{2,1}-u_{1,2}=0,\qquad u_{3,1}-u_{1,3}=0,\qquad u_{4,1}-u_{1,4}=0,\\
&u_{3,2}-u_{2,3}=0,\qquad u_{4,2}-u_{2,4}=0,\qquad u_{4,3}-u_{3,4}=0,\\
&\kappa_1 H+\sum_{r}\kappa_r\frac{\partial H}{\partial u_{i,j}}
+\sum_{s}\kappa_s\frac{\partial^2 H}{\partial u_{k,l}\partial u_{m,n}}+\sum_{r}\kappa_{r+31}u_{i,j}+\kappa_{43}=0,
\end{aligned}
\end{equation}
As done in the previous subsection,  the substitutions
\begin{equation}
\begin{aligned}
&u_1\rightarrow u_1+\alpha_1 x_1+\alpha_2 x_2+\alpha_3 x_3+\alpha_4 x_4, \\
&u_2\rightarrow u_2+\alpha_2 x_1+\alpha_5 x_2+\alpha_6 x_3+\alpha_7 x_4, \\
&u_3\rightarrow u_3+\alpha_3 x_1+\alpha_6 x_2+\alpha_8 x_3+\alpha_9 x_4,\\
&u_4\rightarrow u_4+\alpha_4 x_1+\alpha_7 x_2+\alpha_9 x_3+\alpha_{10} x_4,\\ 
\end{aligned}
\end{equation}
provided that $\kappa_{43}$ can be suitably expressed in terms of the remaining coefficients and of the constants
$\alpha_i$ $(i=1,\ldots,10)$, we have a homogeneous system like (\ref{monge3+1_equiv}) where we can assume 
$\kappa_{43}=0$.

This nonlinear system admits the Lie symmetries spanned by the operators
\begin{equation}
\begin{aligned}
&\Xi_1=\frac{\partial}{\partial x_1},\qquad
\Xi_2=\frac{\partial}{\partial x_2},\qquad \Xi_3=\frac{\partial}{\partial x_3},\qquad \Xi_4=\frac{\partial}{\partial x_4},\\
&\Xi_5=\left(x_1-f_{;1}\right)\frac{\partial}{\partial x_1}+\left(x_2-f_{;2}\right)\frac{\partial}{\partial x_2}+\left(x_3-f_{;3}\right)\frac{\partial}{\partial x_3}+\left(x_4-f_{;4}\right)\frac{\partial}{\partial x_4},
\end{aligned}
\end{equation}
where $f(u_1,u_2,u_3,u_4)$ is a smooth arbitrary function of the indicated arguments, provided that
$\kappa_i$ $(i=1,\ldots,32)$ must be expressed suitably in terms of $\kappa_j$ $(j=33,\ldots,42)$ and $f(u_1,u_2,u_3,u_4)$:
\begin{align*}
&\kappa_1+H_f^{11}\kappa_{33}+H_f^{12}\kappa_{34}+H_f^{13}\kappa_{35}+H_f^{14}\kappa_{36}
+H_f^{22}\kappa_{37}+H_f^{23}\kappa_{38}+H_f^{24}\kappa_{39}\\
&+H_f^{33}\kappa_{40}+H_f^{34}\kappa_{41}+H_f^{44}\kappa_{42}=0,\displaybreak[0]\\
&\kappa_2  + (f_{;34}^2-f_{;33}f_{;44})\kappa_{37}+ (f_{;23}f_{;44}- f_{;24}f_{;34})\kappa_{38}
        + (f_{;24}f_{;33}-f_{;23}f_{;34})\kappa_{39}\\  
 &+ (f_{;24}^2- f_{;22}f_{;44})\kappa_{40}
 + (f_{;22}f_{;34}- f_{;23}f_{;24})\kappa_{41}+(f_{;23}^2- f_{;22}f_{;33})\kappa_{42}=0,\displaybreak[0]\\
&2\kappa_3+ (f_{;33}f_{;44} - f_{;34}^2)\kappa_{34}+(f_{;24}f_{;34}- f_{;23}f_{;44})\kappa_{35}
+ (f_{;23}f_{;34}- f_{;24}f_{;33})\kappa_{36}\\
& + (f_{;14}f_{;34}- f_{;13}f_{;44})\kappa_{38} 
+ (f_{;13}f_{;34}- f_{;14}f_{;33})\kappa_{39}+ 2(f_{;12}f_{;44}-f_{;14}f_{;24})\kappa_{40}\\
&+ (f_{;13}f_{;24}- 2f_{;12}f_{;34} + f_{;14}f_{;23})\kappa_{41}
+2(f_{;12}f_{;33}- f_{;13}f_{;23})\kappa_{42}=0,\displaybreak[0]\\
&2\kappa_4+ (f_{;24}f_{;34}- f_{;23}f_{;44})\kappa_{34}
+ (f_{;22}f_{;44}- f_{;24}^2) 
 + (f_{;23}f_{;24}- f_{;22}f_{;34})\kappa_{36}\\
&+  2(f_{;13}f_{;44} - f_{;14}f_{;34})\kappa_{37} 
+ (f_{;14}f_{;24}- f_{;12}f_{;44})\kappa_{38}\\
&+ (f_{;12}f_{;34}+f_{;14}f_{;23} -2f_{;13}f_{;24})\kappa_{39}+ (f_{;12}f_{;24}- f_{;14}f_{;22})\kappa_{41}\\
 &+ 2(f_{;13}f_{;22}-f_{;12}f_{;23})\kappa_{42}=0,\displaybreak[0]\\
&2\kappa_5+ (f_{;23}f_{;34} - f_{;24}f_{;33})\kappa_{34}
+ (f_{;23}f_{;24}- f_{;22}f_{;34})\kappa_{35}
+ (f_{;22}f_{;33}- f_{;23}^2)\kappa_{36}\\
&+2(f_{;14}f_{;33}-f_{;13}f_{;34})\kappa_{37}
+ (f_{;12}f_{;34}+ f_{;13}f_{;24}- 2f_{;14}f_{;23})\kappa_{38}\\
&+ (f_{;13}f_{;23}- f_{;12}f_{;33})\kappa_{39} 
 + 2(f_{;14}f_{;22}-f_{;12}f_{;24})\kappa_{40}
+(f_{;12}f_{;23}- f_{;13}f_{;22})\kappa_{41}=0,\displaybreak[0]\\
&\kappa_6+ (f_{;34}^2- f_{;33}f_{;44})\kappa_{33}
+ (f_{;13}f_{;44}- f_{;14}f_{;34})\kappa_{35}+ (f_{;14}f_{;33}-f_{;13}f_{;34})\kappa_{36}\\
& + (f_{;14}^2- f_{;11}f_{;44})\kappa_{40}
 + (f_{;11}f_{;34} - f_{;13}f_{;14})\kappa_{41}
  + (f_{;13}^2- f_{;11}f_{;33})\kappa_{42}=0,\displaybreak[0]\\
& 2\kappa_7+ 2(f_{;23}f_{;44}-f_{;24}f_{;34})\kappa_{33} 
 + (f_{;14}f_{;34}-f_{;13}f_{;44})\kappa_{34}
 + (f_{;14}f_{;24}- f_{;12}f_{;44})\kappa_{35}\\
& + (f_{;12}f_{;34}+ f_{;13}f_{;24}- 2f_{;14}f_{;23})\kappa_{36} 
 + (f_{;11}f_{;44}- f_{;14}^2)\kappa_{38}\\
&+ (f_{;13}f_{;14} - f_{;11}f_{;34})\kappa_{39}
 + (f_{;12}f_{;14}- f_{;11}f_{;24})\kappa_{41}  
+ 2(f_{;11}f_{;23}- f_{;12}f_{;13})\kappa_{42}=0,\displaybreak[0]\\
& 2\kappa_8+ 2(f_{;24}f_{;33}- f_{;23}f_{;34})\kappa_{33}
+(f_{;13}f_{;34}- f_{;14}f_{;33})\kappa_{34}\\
& + ( f_{;12}f_{;34}+f_{;14}f_{;23}- 2f_{;13}f_{;24})\kappa_{35} 
 + (f_{;13}f_{;23}- f_{;12}f_{;33})\kappa_{36}\\
& + (f_{;13}f_{;14}- f_{;11}f_{;34})\kappa_{38}
 + (f_{;11}f_{;33}- f_{;13}^2)\kappa_{39} 
 + 2(f_{;11}f_{;24}-f_{;12}f_{;14})\kappa_{40}\\
& + (f_{;12}f_{;13} - f_{;11}f_{;23})\kappa_{41}=0, \displaybreak[0]\\
&\kappa_9+ (f_{;24}^2- f_{;22}f_{;44})\kappa_{33}
+ (f_{;12}f_{;44}- f_{;14}f_{;24})\kappa_{34}
+ (f_{;14}f_{;22}- f_{;12}f_{;24})\kappa_{36}\\
&+ (f_{;14}^2- f_{;11}f_{;44})\kappa_{37}
 + (f_{;11}f_{;24}- f_{;12}f_{;14})\kappa_{39} 
 + (f_{;12}^2- f_{;11}f_{;22})\kappa_{42}=0,\displaybreak[0]\\
 &2\kappa_{10} + 2(f_{;22}f_{;34}- f_{;23}f_{;24})\kappa_{33}
+ (f_{;13}f_{;24} + f_{;14}f_{;23}- 2f_{;12}f_{;34})\kappa_{34}\\
&+ (f_{;12}f_{;24}- f_{;14}f_{;22})\kappa_{35} 
+ (f_{;12}f_{;23}- f_{;13}f_{;22})\kappa_{36}
+ 2(f_{;11}f_{;34}-f_{;13}f_{;14})\kappa_{37}\\
&+ (f_{;12}f_{;14}- f_{;11}f_{;24})\kappa_{38} 
 + (f_{;12}f_{;13}-f_{;11}f_{;23})\kappa_{39} 
+(f_{;11}f_{;22} - f_{;12}^2)\kappa_{41}=0,\displaybreak[0]\\
& \kappa_{11}+(f_{;23}^2- f_{;22}f_{;33})\kappa_{33} 
+ (f_{;12}f_{;33}- f_{;13}f_{;23})\kappa_{34}
+ (f_{;13}f_{;22}- f_{;12}f_{;23})\kappa_{35} \\
&+ (f_{;13}^2- f_{;11}f_{;33})\kappa_{37}
+ (f_{;11}f_{;23}- f_{;12}f_{;13})\kappa_{38}
+ (f_{;12}^2- f_{;11}f_{;22})\kappa_{40}=0,\displaybreak[0]\\
& 2\kappa_{12}- f_{;44}\kappa_{40}+ f_{;34}\kappa_{41} - f_{;33}\kappa_{42}=0,\displaybreak[0]\\
& 2\kappa_{13}  + f_{;44}\kappa_{38}- f_{;34}\kappa_{39}
 - f_{;24}\kappa_{41}+2f_{;23}\kappa_{42}=0,\displaybreak[0]\\
 &2\kappa_{14} - f_{;34}\kappa_{38}+ f_{;33}\kappa_{39}
 + 2f_{;24}\kappa_{40} - f_{;23}\kappa_{41} =0,\displaybreak[0]\\
& 2\kappa_{15}+ f_{;44}\kappa_{35} 
 - f_{;34}\kappa_{36} - f_{;14}\kappa_{41} + 2f_{;13}\kappa_{42}=0,\displaybreak[0]\\
 &2\kappa_{16} - f_{;34}\kappa_{35} + f_{;33}\kappa_{36}
  + 2f_{;14}\kappa_{40} - f_{;13}\kappa_{41}=0,\displaybreak[0]\\
& 2\kappa_{17}+ f_{;34}\kappa_{34}- f_{;23}\kappa_{36}
 - f_{;14}\kappa_{38} +f_{;12}\kappa_{41}=0,\displaybreak[0]\\
&2\kappa_{18} - f_{;44}\kappa_{37} + f_{;24}\kappa_{39} - f_{;22}\kappa_{42}=0,\displaybreak[0]\\
&2\kappa_{19}+ 2f_{;34}\kappa_{37}- f_{;24}\kappa_{38}- f_{;23}\kappa_{39}
  +f_{;22}\kappa_{41}=0,\displaybreak[0]\\
 &2\kappa_{20}+ f_{;44}\kappa_{34}- f_{;24}\kappa_{36} - f_{;14}\kappa_{39}
+2f_{;12}\kappa_{42}=0,\displaybreak[0]\\
& 2\kappa_{21} + f_{;24}\kappa_{35} - f_{;23}\kappa_{36}
 - f_{;14}\kappa_{38}+f_{;13}\kappa_{39}=0,\displaybreak[0]\\
&2\kappa_{22}- f_{;24}\kappa_{34} 
 + f_{;22}\kappa_{36}+ 2f_{;14}\kappa_{37}- f_{;12}\kappa_{39}=0,\displaybreak[0]\\
& 2\kappa_{23}- f_{;33}\kappa_{37} + f_{;23}\kappa_{38}  - f_{;22}\kappa_{40}=0,\displaybreak[0]\\
&2\kappa_{25}+ f_{;33}\kappa_{34}- f_{;23}\kappa_{35} - f_{;13}\kappa_{38} 
+2f_{;12}\kappa_{40} =0,\displaybreak[0]\\
&2\kappa_{26}- f_{;23}\kappa_{34}  + f_{;22}\kappa_{35} + 2f_{;13}\kappa_{37}- f_{;12}\kappa_{38}
 =0,\displaybreak[0]\\
& 2\kappa_{27} - f_{;44}\kappa_{33} + f_{;14}\kappa_{36} - f_{;11}\kappa_{42}=0,\displaybreak[0]\\
&2\kappa_{28} + 2f_{;34}\kappa_{33}- f_{;14}\kappa_{35}
 - f_{;13}\kappa_{36} +f_{;11}\kappa_{41}=0,\displaybreak[0]\\
& 2\kappa_{29}+ 2f_{;24}\kappa_{33}- f_{;14}\kappa_{34}
 - f_{;12}\kappa_{36} +f_{;11}\kappa_{39}=0,\displaybreak[0]\\
& 2\kappa_{30} - f_{;33}\kappa_{33}+ f_{;13}\kappa_{35}  - f_{;11}\kappa_{40}=0,\displaybreak[0]\\
&2\kappa_{31}+ 2f_{;23}\kappa_{33}- f_{;13}\kappa_{34} - f_{;12}\kappa_{35}
   +f_{;11}\kappa_{38}=0,\displaybreak[0]\\
&2\kappa_{32}  - f_{;22}\kappa_{33}+ f_{;12}\kappa_{34} - f_{;11}\kappa_{37}=0.
\end{align*}
Also in this case, the previous conditions place severe restrictions on the coefficients of
system (\ref{monge3+1_equiv}) since imply that the functions $\kappa_i$ $(i=1,\ldots,32)$ have to be expressed
in terms of the coefficients $\kappa_i$ $(i=33,\ldots,42)$ and the function $f$. 

Due to
\begin{equation}
\left[\Xi_i,\Xi_j\right]=0,\qquad \left[\Xi_i,\Xi_5\right]=\Xi_5,
\qquad (i,j=1,\ldots,4),
\end{equation}
we introduce the new variables
\begin{equation}
\begin{array}{llll}
\label{newvars31}
z_1=x_1-f_{;1},\quad &z_2=x_2-f_{;2},\quad & z_3= x_3-f_{;3},\quad & z_4= x_4-f_{;4},\\ 
w_1=u_1,\quad  &w_2=u_2,\quad & w_3=u_3,\quad & w_4=u_4,
\end{array}
\end{equation}
and the generators of the point symmetries write as
\begin{equation}
\begin{aligned}
&\Xi_1=\frac{\partial}{\partial z_1}, \qquad \Xi_2=\frac{\partial}{\partial z_2}, \qquad
\Xi_3=\frac{\partial}{\partial z_3},\qquad
\Xi_4=\frac{\partial}{\partial z_4},\\
&\Xi_5=z_1\frac{\partial}{\partial z_1}+z_2\frac{\partial}{\partial z_2}
+z_3\frac{\partial}{\partial z_3}+z_4\frac{\partial}{\partial z_4}.
\end{aligned} 
\end{equation}

In terms of the new variables (\ref{newvars31}), the nonlinear system (\ref{monge3+1_equiv}) 
assumes the form of an autonomous and homogeneous quasilinear system,
\begin{equation}
\begin{aligned}
&w_{2,1}-w_{1,2}=0,\qquad w_{3,1}-w_{1,3}=0,\qquad w_{4,1}-w_{1,4}=0,\\
&w_{3,2}-w_{2,3}=0,\qquad w_{4,2}-w_{2,4}=0,\qquad w_{4,3}-w_{3,4}=0,\\
&\kappa_{33} w_{1,1} + \kappa_{34} w_{1,2} + \kappa_{35} w_{1,3} + \kappa_{36} w_{1,4}
+\kappa_{37} w_{2,2}+ \kappa_{38} w_{2,3}+\kappa_{39} w_{2,4}\\
&+\kappa_{40} w_{3,3}+\kappa_{41} w_{3,4}+\kappa_{42} w_{4,4}=0,
\end{aligned}
\end{equation}
where $\kappa_i=\kappa_i(w_1,w_2,w_3,w_4)$ $(i=33,\ldots,42)$.

\subsection*{Acknowledgements}
Work supported by ``Gruppo Nazionale per la Fisica Matematica'' 
(G.N.F.M.) of the ``Istituto Nazionale di Alta Matematica'' (I.N.d.A.M.).

\end{document}